\documentclass[]{spie}

\usepackage{amsmath,amsfonts,amssymb}
\usepackage{graphicx}
\usepackage[colorlinks=true, allcolors=blue]{hyperref}
\usepackage{comment}
\usepackage{wrapfig}
\usepackage{threeparttable}
\usepackage{floatrow}
\usepackage{color, soul, subfigure}

\newcommand{\aprx}{\mbox{$\ensuremath{\sim}$}}

\title{A Self-Learning Neural Network Approach for RFI Detection and Removal in Radio Astronomy}

\author[a]{Benjamin R.~B. Saliwanchik}
\author[a]{An\v{z}e Slosar}

\affil[a]{Department of Physics, Brookhaven National Laboratory, Upton, NY, USA}

\authorinfo{Further author information: (Send correspondence to B.R.B.S.)\\ E-mail: bsaliwanc@bnl.gov}

\begin{document} 
\maketitle

\begin{abstract}
We present a novel neural network (NN) method for the detection and removal of Radio Frequency Interference (RFI) from the raw digitized signal in the signal processing chain of a typical radio astronomy experiment. The main advantage of our method is that it does not require a training set. Instead, our method relies on the fact that the true signal of interest coming from astronomical sources is thermal and therefore described as a Gaussian random process, which cannot be compressed.  We employ a variational encoder/decoder network to find the compressible information in the  datastream that can explain the most variance with the fewest degrees of freedom. We demonstrate it on a set of toy problems and stored ringbuffers from the Baryon Mapping eXperiment (BMX) prototype. 
We find that the RFI subtraction is effective at cleaning simulated timestreams: while we find that the power spectra of the RFI-cleaned timestreams output by the NN suffer from extra signal consistent with additive noise, we find that it is generally around percent level across the band and sub 10 percent in contaminated spectral channels even when RFI power is an order of magnitude larger than the signal. We discuss advantages and limitations of this method and possible implementation in the front-end of future radio experiments.
\end{abstract}

\section{Introduction}
\label{sec:intro}

Radio Frequency Interference (RFI) is a significant issue for the operation of any radio frequency telescope, and especially for 21\,cm intensity mapping experiments, which need to achieve exceptionally high mapping sensitivities and dynamic ranges in order to achieve their scientific goals.

The method of 21~cm intensity mapping is a promising new approach to measuring the Baryon Acoustic Oscillation (BAO) structure using the hyperfine splitting line of neutral hydrogen, and which has the potential to complement existing methods of measuring the Large Scale Structure of the universe \cite{Chang:2007xk}. A number of experiments are currently operating, or are under development to attempt to measure the 21~cm cosmological signal, including CHIME \cite{bandura14}, HIRAX \cite{newburgh16}, and Tianlai\cite{chen12}, and plans are already being made for next-generation arrays such as CHORD\cite{vanderlinde19} and PUMA\cite{slosar19}, which will incorporate orders of magnitude more elements, to achieve correspondingly higher sensitivities. 

The chief limitation of these experiments is the extremely low level of the expected signal, of order 0.1mK, and the extremely bright foreground radiation from sources such as the Milky Way galaxy. This low amplitude signal in turn results in the necessity of extremely large interferometric arrays. Current generation arrays consist of hundreds or one thousand dish elements, with diameters typically 6-14~m. It also means that reducing and removing RFI is critical for the operation of such instruments, if they are to achieve the sensitivities necessary to detect the cosmological 21~cm signal. Site selection and instrument design will be critical components of achieving low RFI contamination, by minimizing the local RFI present, and shielding the instrument from local RFI sources. Radio instruments are usually located at remote sites to avoid man-made interference, and ideally with geographical features that shield the telescope site. Unintentional radiators on the observing site can also be restricted, such as by prohibiting various electronic items, and by surveying for and removing accidental sources such as faulty electronics or wiring. Additionally, instruments are designed with ground shields or other structures which are designed to reduce the amount of local interference that can enter the instrument, and sensitive electronics (or electronics which may produce RFI) can be shielded to prevent pickup (or radiation). However, it is not possible to completely avoid or suppress all sources of RFI. Of particular concern are in-band sources that are not local, but which cannot be blocked from entering instruments, such as satellites, TV broadcast stations, cell phones, and air traffic control radar and communications. There are numerous such sources, and we can certainly expect the number to increase in future. Given the impossibility of avoiding all RFI sources, it is necessary to also develop methods of recognizing and removing RFI from instrument timestreams.

The practical methods in use for RFI detection work with spectral data, i.e. after the raw waveform data has been processed and reduced in volume through the spectrometer. The aim is to isolate bins which are contaminated with RFI. There are three main classes of methods for RFI detection from spectra in the literature. The first category is linear algebraic methods such as Singular Value Decomposition (SVD)\cite{offringa10} and Principle Component Analysis (PCA)\cite{zhao13}, which are best at recognizing repeated patterns in time and frequency space, but do not handle irregular signals well. Second are threshold based methods, which detect pixels above some threshold, typically after high-pass filtering \cite{baan04, offringa10}. The third are more ``traditional'' supervised machine learning methods such as Gaussian mixture models and K-nearest neighbors models \cite{wolfaardt16}. More recently, neural networks have become a promising new category of methods which may significantly improve our ability to recognize and remove RFI. Numerous machine learning techniques have been proposed for the problem of RFI identification and removal, including random forest classifiers \cite{mosiane17}, long short-term memory (LSTM) \cite{czech18}, You Only Look Once (YOLO) \cite{ghanney20}, and convolutional neural networks (CNN) \cite{czech18, akeret17, sadr20}. A limitation of these methods is that they are supervised learning techniques, and require accurate models of all forms of RFI which will be encountered in order to be trained. A more desirable, and potentially more adaptable, class of methods would be unsupervised learning methods, which do not require prior knowledge of the morphology of RFI signals. Several unsupervised learning methods have also been tested for RFI removal, including Convolutional Auto-Encoders (CAE) \cite{ghanney20}, and Generative Adversarial Networks (GAN) \cite{vos19}. It should be noted that machine learning methods for RFI flagging or removal are still in the early stages of development. To the best of our knowledge, no such method has been used to date as the final production method of RFI processing on published astrophysical data. The methods presented here, and the others referred to above, are experimental methods.

The insight which led to the development of the method described here is that time ordered data from thermal astrophysical sources can in general be distinguished from RFI by the fact that the amplitudes and phases of the astrophysical signals follow a Gaussian random distribution. Most RFI detection methods are applied to spectra. However, the process of compressing the waveforms into spectra is lossy: not only is all the phase information gone, but there has also been significant averaging. For a pure signal from the radio sky, this compression is harmless: radio signals from the radio sky are thermal in nature and thus follow statistics of stationary Gaussian random fields. Such fields are statistically completely described by their power spectra. On the other hand, RFI signals are either meant to communicate information and are thus designed to be separable from Gaussian noise, or alternatively, they are accidental emissions that are localized in time and thus break the stationarity of the sky signal. In either case, the purpose of RFI removal is to remove any non-Gaussianity in the signal as any stationary thermal RFI is \textit{a priori} inseparable from the true sky signal.

Of course, there is no such thing as a general non-Gaussianity. As a thought example, imagine a signal composed of a sequence of $N$ Gaussian variates that repeats itself. When seeing the sequence for the very first time, we theoretically cannot distinguish it from the Gaussian signal, however a general enough method will work out that it repeats itself and isolate it. Therefore, the hunt for non-Gaussian signals hidden in Gaussian noise without prior information must inherently be a statistical process. After seeing a sufficient number of data examples, a method can learn about the statistical properties of the non-Gaussianities present and remove them from future data. Here we propose one such  method, which makes use of the deep convolutional neural network architecture to isolate RFI in an unsupervised manner. We note from the beginning that any such method will isolate all non-stationary signals, including those from astronomically interesting transients, such as fast enough fast radio bursts (FRBs). This method was designed with next generation 21~cm arrays in mind, but is in principle applicable to any radio frequency instrument. 

Of course, any method that operates on the raw data-stream must be both more efficient as well a computationally considerably more intensive. However, with the advent of novel Radio Frequency System-on-Chip (RFSoC) technologies, it is entirely feasible to perform analysis and potential RFI removal from the datastream before it hits the spectrometer code. In this work we focus on a toy demonstration of the method using code running on a PC. This illustration is to motivate development of this code for a future production scale system; we discuss full-scale feasibility in more detail in the conclusions section.

\section{Method}

Convolutional neural networks excel at image recognition, and usually function by  repeatedly applying convolutions to the input image, each followed by an activation function and downsampling layer. These repeated convolutions build up a conceptual hierarchy of the contents of the image, while the downsampling reduces the image to the most salient features. This structure is loosely analogous to biological visual processing systems, where layers of neurons first extract small scale features such as edges, and then combine the small scale features into more complex objects and patterns. Akeret et al. \cite{akeret17} pioneered a modified CNN architecture that takes this standard downsampling CNN architecture, and appends an inverted series of operations which perform additional convolutional layers, and upsample. They call this architecture the ``U-Net" due to the symmetric downsampling and upsampling layers, which they term ``encoder'' and ``decoder'' layers. The advantage of this approach is that the complex features detected with the encoder layer can then be upsampled with the decoder layer, to produce an output of the same dimensionality as the input data. They then use a $1 \times 1$ convolution to map from the last layer to a binary decision, i.e. is the pixel RFI contaminated? A pixel-wise soft max layer then gives the probability of each pixel belonging to the decided category. RFI mitigation is performed by inputting two dimensional time ordered data (frequency vs. time), and using the output probabilities to threshold mask the input based on the probability of RFI contamination.

Our NN architecture is based on that of the U-Net, but crucially, it does not operate on spectral data, but on raw timestreams and our objective is not simply to apply a threshold mask, but to characterize the structure of RFI contamination, and subtract it from the time ordered data to recover the underlying uncontaminated sky data. 

Let's start by splitting the observed timestream $S$ into its Gaussian and non-Gaussian components:
\begin{equation}
    S = T_g + T_{ng}.
\end{equation}
The Gaussian component $T_g$ is statistically fully characterized by its power spectrum. If we renormalize the signal in Fourier space so that the power spectrum of $T_g$ is white, then $T_g$ is a set of uncorrelated Gaussian variates. $T_{ng}$ is signal which cannot be described as $T_g$ and describes our RFI contamination.

Consider a neural network that takes an input signal and returns an estimate of the non-Gaussian signal, i.e.  $N_N(S) = T_{ng}$ in an ideal case. If we had an example of $T_{ng}$ we could simply train the neural network to recover it. Assuming we knew $T_g$, we could train it by demanding that $S-N_N(S)$ recovers $T_g$.  Unfortunately, we do not \textit{a priori} know either; all we have is the sum of two signals.

However, assuming we know the power spectrum of $T_g$ (or assuming it can be approximated with the power spectrum of $S$), we can generate a \emph{known} Gaussian signal $T_g'$. This can be accomplished by Fourier transforming $S$ from the time domain to frequency domain, multiplying each element by a random phase, and Fourier transforming back into a time ordered series, $T_g'$, which has approximately the same power spectrum as $T_g$, but is completely known. We can then use $T_g'$ to create a new signal, $S'$ using a weight $0<\lambda<1$, as follows
\begin{equation}
    S' = \sqrt{1-\lambda^2} S + \lambda T'_{g}
\end{equation}
and then train the network to recover $S'-N_N(S') = \lambda T'_{g}$. In practice we do this by minimizing the $L_2$ distance or sum of square residuals
\begin{equation}
    {\rm Loss} = \sum_{\rm samples} (S'-N_N(S')-\lambda T'_{g})^2
\end{equation}

\begin{figure}[H]
    \centering
    \includegraphics[width=\textwidth]{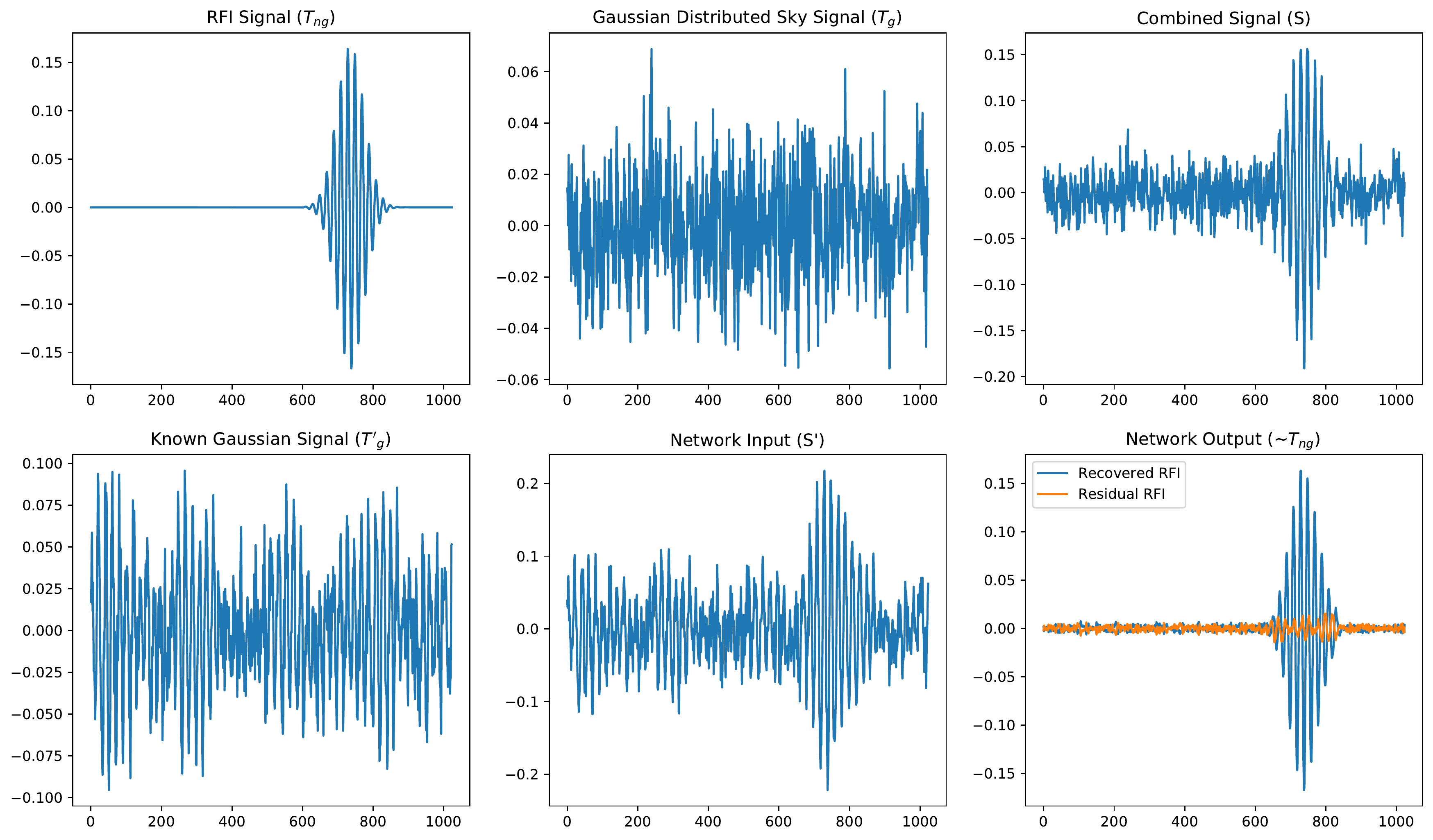}
    \caption{Input and output timestreams from our neural network. Horizontal axis is sample number, vertical is arbitrary amplitude units. The top left panel shows the injected RFI signal, $T_{ng}$, which we attempt to characterize, modeled as a sinusoidal oscillation within a Gaussian amplitude envelope. Top middle shows the modeled sky signal, $T_g$ composed of a timestream of Gaussian random distributed amplitudes and phases, multiplied by a noise power spectrum, $P_k$. Top right shows the combination of these two signals, $S$, the signal which would be observed by a radio telescope. This simulated RFI event has a signal-to-noise ratio of \aprx4, which is common for radio instruments. Bottom left shows $T'_g$, the known Gaussian signal which we combine with $S$ in order to train our NN. Bottom middle is $S'$, the normalized sum of $S$ and $T'_g$ which is input to the NN. Bottom right in blue is the output of the NN, which clearly isolates the non-Gaussian RFI component of the input signal, including position, frequency, phase, and amplitude structure. Overplotted in orange is the residual RFI which is not reconstructed by the NN. The accuracy of the recovered signal is explored further in Section \ref{sec:optim}.}
    \label{fig:NN}
\end{figure}

Let's take a step back to see what is happening here. Signals $T_g$ and $T'_g$ are statistically indistinguishable. The neural network is mathematically incapable of separating the two, so at best it can hope to learn to pick out the non-Gaussian part (so that $N_n(S')\propto T_{ng}$). In this case, a properly trained network will just result in a irreducible loss function corresponding to $\sim T_{g}^2$. Figure \ref{fig:NN} shows an example of the input and output signals to our NN, demonstrating that this process is effective at recovering an injected non-Gaussian RFI contaminant signal. Section \ref{sec:pwr_spec} below discusses our verification that the input sky signals are not significantly altered by this RFI removal process.

In principle, this works for any sufficiently large $\lambda$. For very small values of $\lambda$, there is also a pathological solution, where $N_n(S) = S$, and the network instead of paying an irreducible loss corresponding to $\sim T_{g}^2$, instead lives with $\sim T_{g}^{\prime \, 2}$. To prevent this from happening, we employ the ``U-net" approach.  The downsampling and upsampling process of our NN forces the data through a minimum dimensional ``chokepoint'', forcing the output of the network to describe as much of the input signal as possible, with a minimum number of free parameters. The sky signal cannot be described by fewer than the original number of parameters (the number of frames in the input timestream), because it is incompressible. However, the RFI signal can be compressed, and thus passes through the encoder-decoder network. In this case, one can train with $\lambda$ that is as small as possible, and in fact we have found that $\lambda=0$ works optimally in our toy-example.  In this example, the loss function is simply
\begin{equation}
    {\rm Loss} = \sum_{\rm samples} (S-N_N(S))^2.
\end{equation}
In other words, we shove the raw data stream through the encoder-decoder network and train it to isolate whatever will decrease the variance the most. A compressible non-Gaussian RFI can consistently decrease power in average. In fact, as described later we find that the form $\lambda=0$ works best in our case and is capable of finding isolated bursts of power. 

The RFI removal would then replace the input signal packet $S$ with its RFI-cleaned version $S-N_N(S)$. The most attractive feature of this approach is that no additional information is required for training, and the NN can be trained unsupervised on raw observed timestreams. In the subsequent section we describe how we implemented and tested the method described here in more detail. 

\section{Optimization of Neural Network}
\label{sec:optim}

Our neural network (NN) is implemented in Python, using the PyTorch library. The architecture consists of an encoder and decoder network, each with the same number of downsampling or upsampling layers. In the baseline version we operate on timestreams with 1024 data frames, and the network consists of three layers on each side. The encoder consists of three linear layers, with Leaky Rectified Linear Units (LeakyReLU) as activation layers after each of the first two layers, with a negative slope of 0.02. These activation layers are necessary for introducing non-linear behavior into the response of the NN. Without them, it functions analogously to a linear algebraic filter. Additionally, a dropout layer is included after the second LeakyReLU activation layer to prevent overfitting. The dropout layer has an element dropout probability of 0.2. The first two linear layers downsample by a factor of two, first to 512 samples, then to 256, and the final layer downsamples to a minimum dimension of 16 samples. We refer to this minimum dimensional choke point as the ``z-dim''. The decoder network has three layers, with LeakyReLU activation layers in between, and upsamples first from 16 to 256 samples, and then to 512, and finally to the original input size of 1024 samples. 

We train the NN for 30 epochs with a training set of 100,000 data timestreams, with a batch size of 32. We use mean squared error as our loss function, and the Adam optimizer, with a learning rate of $\gamma = 0.0002$, and $(\beta_1, \beta_2) = (0.5, 0.999)$. Adam was selected as the most efficient optimizer, and the learning rate and beta values were further tuned to improve the fitting rate for our test data. The effects of varying the number of training epochs and data sets, along with other training parameters, are described in Section \ref{sec:training} below.

For the optimization studies discussed in this section, we train the NN with artificial data, so that we have a known true model of the sky data and RFI, and can assess how accurately the NN extracts the RFI. This artificial data is constructed from a Gaussian random ``sky signal'', and potentially one or more RFI events. The sky data in constructed in the frequency domain, by generating an array of complex numbers $X_k$ whose real and imaginary parts are sampled from a Gaussian random distribution. This array is then multiplied by a noise power spectrum. The baseline noise power spectrum used is:

\begin{equation}
    P(k) = \alpha \left(1 + e^{-\beta (k - k_0)^2}\right) e^{-k / k_0},
\end{equation}

where $k_0 = 256$, $\beta = 0.0002$, and $\alpha$ is a normalization factor to account for varying the length of the timestream. We also explore using a white noise spectrum in Section \ref{sec:training}. This random complex array, scaled by the noise power spectrum, is then Fourier transformed to produce a synthetic timestream for the Gaussian random distributed sky signal:

\begin{equation}
    T_g(n) = \sum_{k=0}^{K-1} P(k) x_k e^{-2i\pi k n / K}, 
\end{equation}

where $K$ is the number of elements in the frequency domain representation of the sky signal, $x_k$ is the $k^{\mathrm{th}}$ frequency element of the sky signal, and P(k) is the noise power spectrum in the $k^{\mathrm{th}}$ frequency bin. The RFI component of the signal, $T_{ng}$ is modeled in the time domain as a sinusoid multiplied by a Gaussian envelope:

\begin{equation}
    T_{ng} = A_0 \ \mathrm{cos}(\phi + \nu t) \ e^{-(t - t_0)^2 / (2 \sigma^2)},
\end{equation}

where $A_0$ is the amplitude, $\phi$ and $\nu$ are the phase and frequency of the sinusoidal component, and $t_0$ and $\sigma$ are the center and width of the Gaussian envelope, and $t$ is time. The position, amplitude, frequency, phase, and Gaussian envelope width are selected from uniform distributions for each synthetic timestream. We train on 100,000 timestreams, and test on ten. In the sections below, where we vary some parameter of the NN, we use a fixed seed for the components of $T_g$ and $T_{ng}$ so that the results can be accurately compared. Throughout, we parameterize the goodness of fit of the RFI timestream recovered by by the NN with the root mean square (RMS) of the difference between the input and output RFI timestreams. To compare different parameterizations or configurations of the NN we use the average RMS over all ten test cases. The NN is trained on a single NVIDIA GeForce RTX 2060 GPU with 6GB of dedicated video memory, and 32GB of system memory.

In Section \ref{sec:norm} we vary the $\lambda$ normalization factor. In Section \ref{sec:encoder} we vary the dimension and number of steps in the encoder-decoder network. In \ref{sec:timestream} we vary the length of the timestream. In Section \ref{sec:events} we explore multiple RFI events in a single timestream. In Section \ref{sec:training} we use different training parameters. And in Section \ref{sec:rfi_params} we explore different parameterizations of the RFI event. The effects of varying these parameters are summarized in Table \ref{tab:nn_params}.

\subsection{Normalization}
\label{sec:norm}

\begin{figure}[H]
\begin{center}
\subfigure{\includegraphics[width=0.49\textwidth]{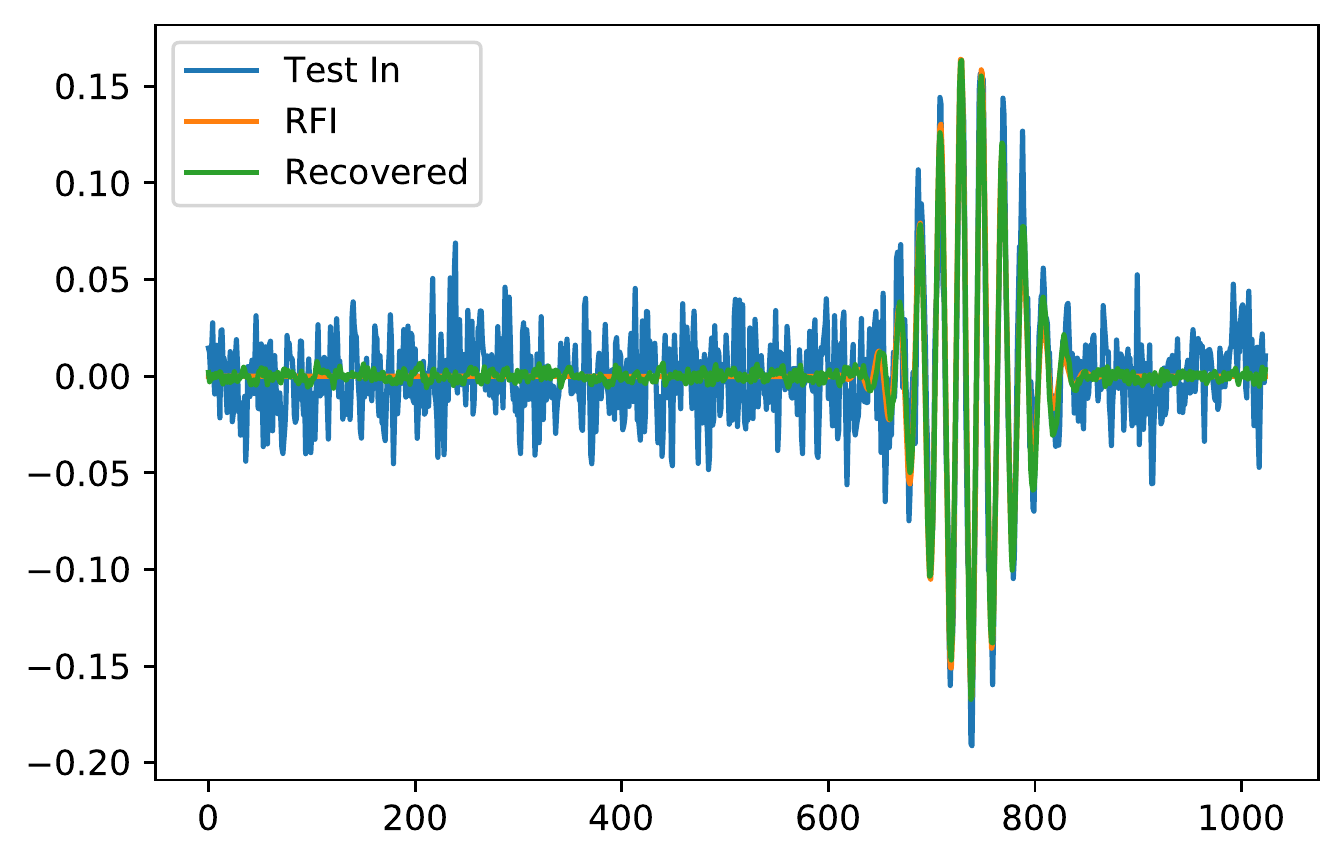}} 
\subfigure{\includegraphics[width=0.49\textwidth]{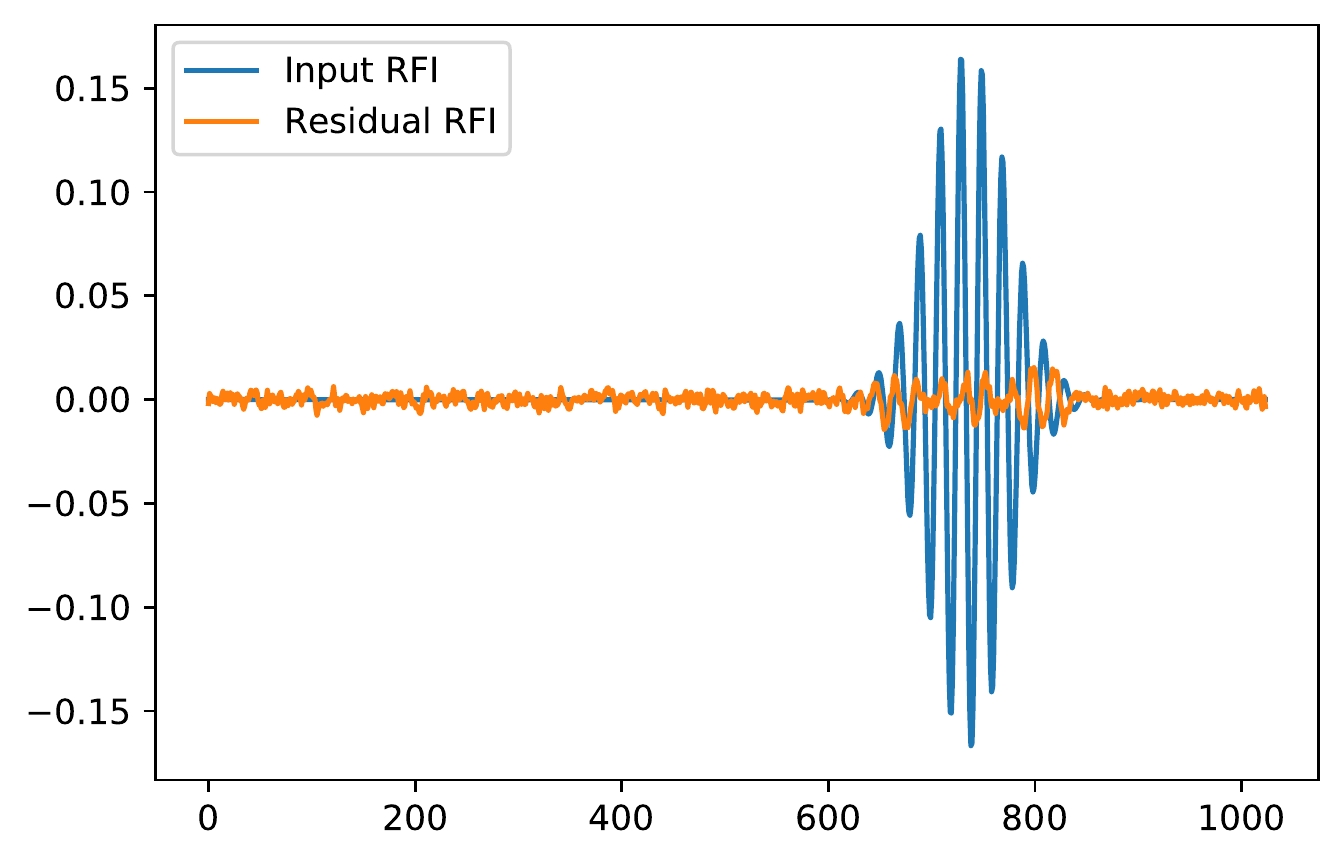}}
\end{center}
\caption[]{(Left) Input and output timestreams from the NN with the normalization factor $\lambda = 0$. Horizontal axis is sample number, vertical is arbitrary amplitude units. The recovered RFI is nearly indistinguishable from the input RFI, demonstrating that the NN functions excellently without the secondary $T'_g$ Gaussian signal. The error in the recovered RFI with $\lambda = 0$ is actually marginally lower than with small finite values of $\lambda$ ($2.5\%$ lower than with $\lambda = 0.1$). (Right) Residual RFI signal compared to the input RFI signal, showing that the majority of the input RFI signal has been recovered by the NN. Section \ref{sec:pwr_spec} below discusses verification that the input sky signal is not significantly altered by this RFI removal process.}
\label{fig:rfi_overplot} 
\end{figure}

We test our NN with four different values of the normalization factor, $\lambda$: 1.0, 0.5, 0.1, and 0. In the case of $\lambda = 0$, no $T_g'$ timestream was generated to speed up processing. We find that the accuracy of the recovered RFI varies substantially with the normalization factor, and tends to improve for smaller values of $\lambda$. For $\lambda = [1.0, 0.5, 0.1, 0.0]$ the average recovered RMS error is $E_{avg} = [0.0226, 0.0065, 0.0041, 0.0040]$, respectively. Figure \ref{fig:rfi_overplot} shows an example of input and output timestreams with the NN trained and evaluated with $\lambda$ set equal to zero, showing that the recovered RFI is still excellently matched to the input timestream in position, frequency, phase, and amplitude. The NN in fact performs marginally better with $\lambda=0$ than with small non-zero $\lambda$ on the test cases examined. While the difference is small enough that it may not seem significant ($2.5\%$ improvement over $\lambda = 0.1$), the major advantage of setting $\lambda = 0$ is that it obviates the necessity of generating the extra $T_g'$ timestream, and significantly decreases training and evaluation time, which is crucial for evaluating real-time telescope data. Therefore in the following sections we set $\lambda = 0$ for all test cases. 

\subsection{Encoder and Decoder Parameters}
\label{sec:encoder}

The shape of the neural network is one of the most basic elements, and can have strong effects on its effectiveness. We performed extensive tests varying the number and size of the linear layer steps in our encoder/decoder network, in order to find the optimum shape of the network for extracting RFI. We begin with a fiducial encoder network with three linear layers. The first layer performs a linear transformation on the 1024 element input array with out reducing it in size, the second linear transform reduces the timestream from 1024 elements to 512, and the third reduces it further from 512 to 16 elements. The decoder network then inverts the operations of the encoder network to produce a 1024 element array which models the RFI in the input array.

The first parameter in this network we set out to optimize is the size of the smallest point in the encoder/decoder network, which we refer to as the z-dimension.  We tested five values for the z-dimension, $z_{dim} = [8, 12, 16, 20, 32]$, and find corresponding average recovered RFI errors of $E_{avg} = [0.0053, 0.0040, 0.0040, 0.0043, \\ 0.0052]$. We find a broad minimum in recovered RFI error around our fiducial value of $z_{dim} = 16$, which we retain as our optimal value since powers of two are convenient for Fourier transforms.

The next optimization was of the size of the intermediate layer, which is often referred to as the dimension of the hidden layer, or the hidden dimension. We vary $h_{dim} = [32, 64, 128, 256, 512, 1024]$, holding the size of the first layer fixed at 1024, and $z_{dim}$ at 16. We recover input RFI with error: $E_{avg} = [0.0173, 0.0091, 0.0056, 0.0048, 0.0040, \\ 0.0045]$.

One can also vary the number of layers in the network. It might be possible for example that more slowly compressing the array from 1024 elements down to our minimum value of 16 allows the NN to more efficiently encode the structure of the RFI. Holding the size of the first linear layer fixed at 1024, and $z_{dim}$ at 16, we vary the number of intermediate layers, with each layer halving the size of the array. Here the number of hidden layers is $N_{HL} = [1, 2, 3, 4, 5, 6]$. This corresponds to a smallest hidden dimension (before the fixed $z_{dim}$ choke point) of $h_{dim} = [1024, 512, 256, 128, 64, 32]$. Here we find $E_{avg} = [ 0.0058, 0.0040, 0.0041, 0.0062, 0.0099, 0.0266]$. Our hypothesis that slower compression would be more effective proved false, and instead there was an optimal length of two layers. We believe this is because some initial linear transformations without reducing the size of the input array are necessary for building the conceptual framework of the image contents, but then compressing quickly down to the choke point is more effective at selecting the few most important structures that describe the RFI.

Since large initial layers, and sharper downsampling seem to be more effective, we thought perhaps a maximally sharp downsampling would be most effective. Instead of reducing the input array from the input size of 1024 to an intermediate hidden layer dimension of 512, and then to 16, perhaps we should reduce it immediately from the input dimension to the choke point dimension in one layer. We tested this hypothesis with networks of varying length, in which each hidden layer transforms from 1024 elements to 1024 elements, and then the final layer downsamples from 1024 to 16 elements. For $N_{HL} = [1, 2, 3, 4, 5]$ we find corresponding recovered RFI errors of $E_{avg} = [0.0058, 0.0045, 0.0037, 0.0047, 0.0266]$. Here a network with two hidden layers performs worse than one with an intermediate downsampling to 512 elements, but a three layer network performs markedly better than our fiducial network. From this point we will proceed using the new encoder/decoder network layout with three linear layers with 1024 elements, and a final one with 16 elements.

\subsection{Training Parameters}
\label{sec:training}

We also optimize over changes to various training parameters including the number of training epochs, the size of the training data set, generating new training data at each epoch, shuffling the training sets vs not shuffling, and training with different noise power spectra.

First we vary, the number of epochs for which the training runs. For $N_{epochs} = [15, 20, 30, 40, 60, 90]$ we find $E_{avg} = [0.0043, 0.0040, 0.0037, 0.0042, 0.0043, 0.0050]$. It is slightly unexpected that past a certain point longer training results in increased error in the recovered RFI in the test cases, but this may be due to overtraining the network. After this test, we stay with 30 epochs of training as optimal.

Next, we vary the size of the training data set, from 25 thousand timestreams up to 200 thousand. For $N_{train} = [25k, 50k, 100k, 150k, 200k]$ we find $E_{avg} = [0.0064, 0.0055, 0.0037, 0.0036, 0.0038]$. Here, after about 100 thousand increasing the training data set does not result in significant improvements in performance, and only increases processing time. We stay with our fiducial choice of 100 thousand training timestreams.

Generating a new training data set each epoch is distinctly worse, both in terms of performance and processing time. The average error when generating a new training set each epoch increases to 0.0048 from 0.0037, while the training time increases by 182\%, to 796 s, from 436 s in our baseline case.

We use batches of 32 timestreams for training, but train using the full data set at each epoch. Our standard procedure is to shuffle the training data between batches, which is supposed to improve training efficiency, and prevent over fitting. However, since we evaluate the loss using the full training set, shuffling the data sets between batches should not in principle improve the training. We test this, comparing the error in recovered RFI with and without shuffling the training data sets, and find no statistical difference in the outcome, as expected. However, it is also not any slower to shuffle the data, so we leave shuffling active, as the standard procedure for neural networks.

Another aspect we considered was the optimal sky signal power spectrum with which to train the network. Most likely the optimal training spectrum is the one found in the test data, and since in practice we can train unsupervised on real sky data, this will be trivial. However, it is possible that a white noise spectrum, will be better at training the neural network to detect RFI in the presence of arbitrary sky signals, since it has equal power at all frequencies. We test this possibility, and find that the error in the recovered RFI increases when using a white spectrum for the training sky signal, and testing on our original sky spectrum. Specifically, the error increases by 8\%, from 0.0037 to 0.0040.

\subsection{Varying Timestream Length}
\label{sec:timestream}

The NN is less effective on longer timestreams, though this effect can be somewhat mitigated. For timestreams of length $N_p = 2048$ the average error is $E_{avg} = 0.0044$, and input RFI events are still recovered with reasonable precision. At $N_p = 4096$ most events are not detected with the baseline NN. However, if we set $z_{dim} = 32$, then the input RFI events are once again detected, and the error for this case is $E_{avg} = 0.0051$.

\subsection{Multiple RFI Events}
\label{sec:events}

Our NN has the ability to simultaneously detect multiple RFI events in single timestream. This can be done without explicitly adding additional model parameters (since there are none), though the efficiency is improved by increasing $z_{dim}$ (which is equivalent to adding more model parameters in some sense). For $z_{dim} = 16$ and a number of RFI signals $N_{rfi} = [1, 2, 3]$ our recovered error is $E_{avg} = [0.0037, 0.0122, 0.0289]$. If we increase $z_{dim}$ to 32 for $N_{rfi} = [2, 3]$ we find $E_{avg} = [0.0085, 0.0175]$.

\subsection{RFI Morphology}
\label{sec:rfi_params}

\begin{figure}[H]
\begin{center}
\includegraphics[width=0.49\textwidth]{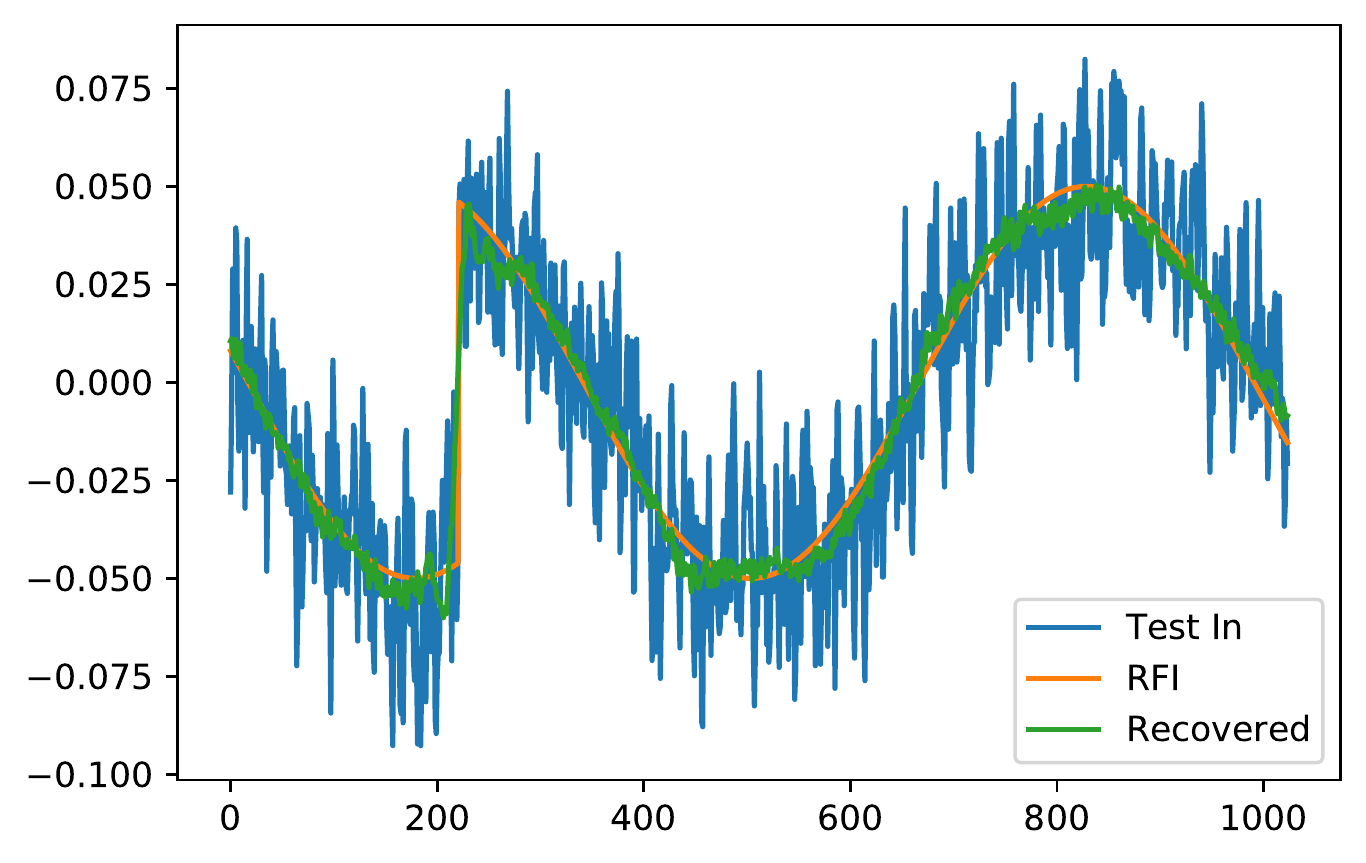}
\includegraphics[width=0.49\textwidth]{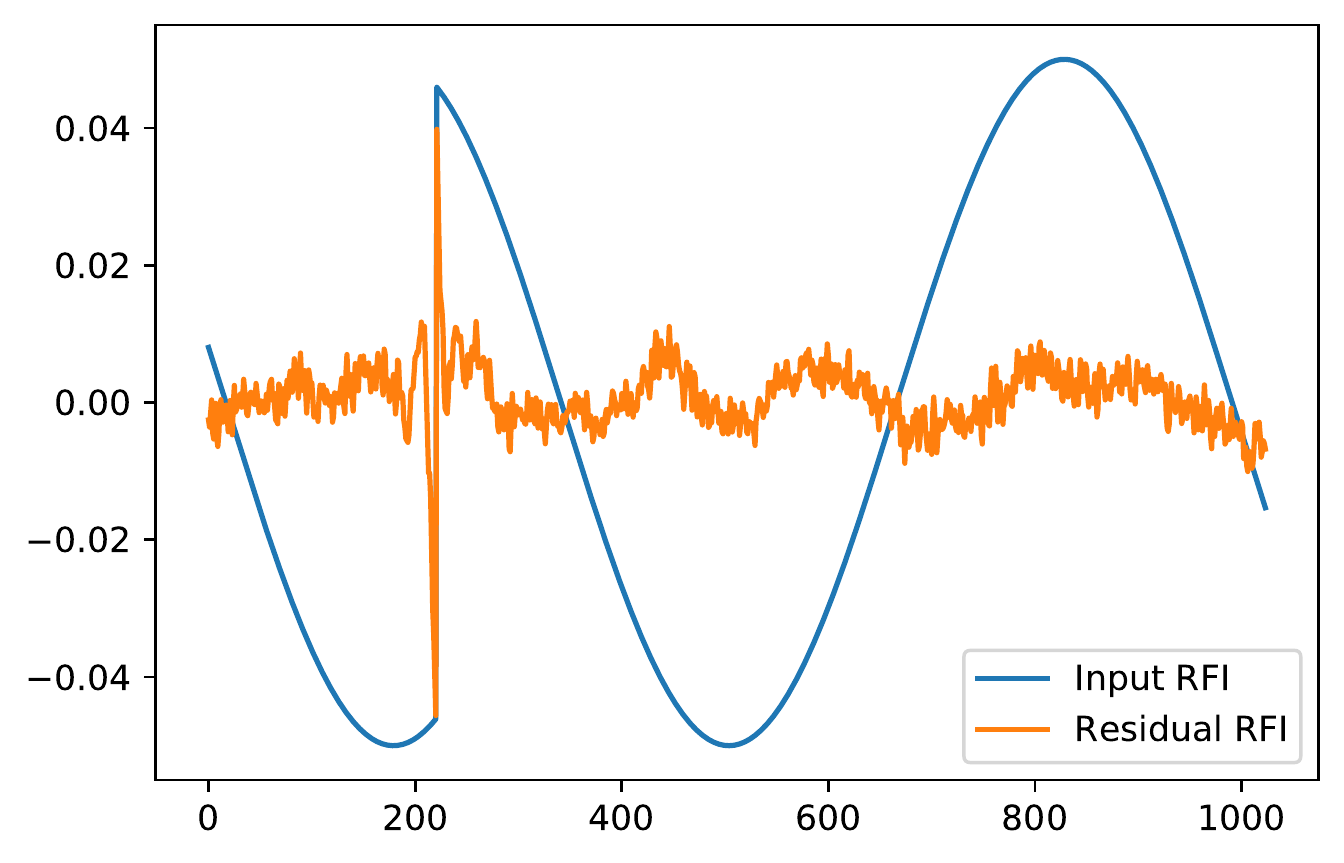}
\end{center}
\caption[]{(Left) Input and output timestreams from the NN with long period sine wave RFI, with a sign-flip discontinuity in the underlying sine wave modeling a phase-shift keying (PSK) digital signal. Horizontal axis is sample number, vertical is arbitrary amplitude units. The NN models this RFI morphology well, despite the discontinuity. (Right) Residual RFI signal compared to the input RFI signal. In most places the majority of the input RFI signal has been recovered by the NN, however because the PSK feature in this RFI model is a discontinuous sign-flip, and the NN output is continuous, the RFI discontinuity produces spikes in the residuals. This feature would be ameliorated in realistic PSK events, which would be smoothed by bandpass filtering, and other effects in either transmission or reception.}
\label{fig:long_period_rfi} 
\end{figure}

There are of course a large number of RFI producing phenomenon to be found in the wild, and a correspondingly large number of RFI morphologies, and we have only tested our NN on a single morphology. We will test our NN on the real-world data from the BMX telescope in Section \ref{sec:bmx}, but first we test one more general test case. So far we have dealt with strongly localized test events, but there might also be RFI events with durations significantly longer than our timestream section length. In this case, rather than a sinusoidal wave within a Gaussian envelope, it might resemble simply a sinusoid of constant amplitude. There might also be discontinuities in the signal, especially if it is digital in origin. Here we will model our RFI signal as a simple sinusoid, potentially multiplied at an arbitrary point by a sign change to model a ``bit-flip'' in a digital signal, to emulate binary phase-shift keying (PSK):

\begin{equation}
T_{ng} = A_0 \ \mathrm{cos}(\phi + \nu t) \left[ -2H(t_0) + 1 \right]
\end{equation}

where $H(t_0)$ is the Heaviside function centered at $t_0$, and the location of the flip feature is randomly selected from the range of the timestream with a uniform distribution. Additionally, we can specify a probability for the sign-flip component so that only some fraction of the timestreams have a flip feature. If $P_{flip} = [0, 0.5, 1.0]$ we find $E_{avg} = [0.0032, 0.0048, 0.0048]$. The NN is very good at fitting long period RFI, but also surprisingly good at fitting sharp features like the sign-flip in this RFI model, even though this feature interrupts the periodic nature of the large scale RFI. This is a very promising feature of this NN, as RFI frequently contains highly discontinuous features. Figure \ref{fig:long_period_rfi} shows an example of the NN recovery of long period sine wave RFI with a sign-flip feature. A caveat here is that this sign-flip is an idealized representation of a PSK signal. In practice a PSK transmitter will usually be bandpass filtered, and other effects either in transmission or reception will further smooth out the otherwise sharp phase shift. Further testing would be required to assess the performance of the NN on smoothed PSK features, but since the NN does not depend on knowing the specific parameterization of the RFI signal, and detects long period RFI without PSK features (which can be viewed as a maximally smoothed phase transition), we expect the performance of the NN with smoothed PSK features to be comparable.

\begin{table*}
\begin{center}
{
\begin{tabular}{c|c|c|c|c|c|c|c}
\multicolumn{8}{c}{Neural Network Parameter Optimization} \\
\hline \hline
 \multicolumn{1}{c|}{NN Parameter} & \multicolumn{1}{c|}{Description} & \multicolumn{6}{|c}{Parameter Values and Errors} \\
\hline
 $\lambda$ & Normalization factor & 1.0 & 0.5 & 0.1 & 0 \\

  & & 0.0226 & 0.0065 & 0.0041 & 0.0040  \\
\hline

 $z_{dim}$ & Minimum layer dimension & 8 & 12 & 16 & 20 & 32 \\

 & & 0.0053 & 0.0040 & 0.0040 & 0.0043 & 0.0052 \\
\hline 

 $h_{dim}$ & Hidden layer dimension & 32 & 64 & 128 & 256 & 512 & 1024 \\

  & & 0.0173 & 0.0091 & 0.0056 & 0.0048 & 0.0040 & 0.0045 \\
\hline 

 $N_{HL}$ & Number of hidden layers & 1 & 2 & 3 & 4 & 5 & 6 \\

  & (decreasing $h_{dim}$) & 0.0058 & 0.0040 & 0.0041 & 0.0062 & 0.0099 & 0.0266 \\
\hline 

 $N_{HL}$ & Number of hidden layers & 1 & 2 & 3 & 4 & 5  \\

  & (constant $h_{dim}$) & 0.0058 & 0.0045 & 0.0037 & 0.0047 & 0.0266 \\
\hline 

 $N_{epochs}$ & Number of training epochs & 15 & 20 & 30 & 40 & 60 & 90  \\

  &  & 0.0043 & 0.0040 & 0.0037 & 0.0042 & 0.0043 & 0.0050 \\
\hline 

 $N_{train}$ & Number of training data sets & 25k & 50k & 100k & 150k & 200k  \\

  &  & 0.0064 & 0.0055 & 0.0037 & 0.0036 & 0.0038 \\
\hline 

 $N_{p}$ & Number of samples & 1024 & 2048 & 4096 \\

  & per timestream & 0.0037 & 0.0044 & 0.0051* \\
\hline 

 $N_{rfi}$ & Number of RFI events & 1 & 2 & 3 \\

  & per timestream & 0.0037 & 0.0122 & 0.0289 \\
\hline 
\hline
\end{tabular}}
\caption{
NN parameters and corresponding average errors, $E_{avg}$, in recovered RFI timestreams. The performance of the NN for RFI characterization can be significantly enhanced by judicious choices for various parameters of the NN architecture. We performed extensive testing to marginalize over these architectural parameters, eventually settling on an encoder-decoder network with three hidden linear transformation layers (the first two with constant dimension of 1024) and a minimum dimension (in the third layer) of 16. *RFI was only reliably detected at $N_p = 4096$ if $z_{dim}$ was increased to 32. The previous two values are with $z_{dim} =  16$.
}
\label{tab:nn_params}
\end{center}
\end{table*}

\subsection{Power Spectrum Recovery}
\label{sec:pwr_spec}

\begin{figure}[H]
\begin{center}
\subfigure{\includegraphics[width=0.49\textwidth]{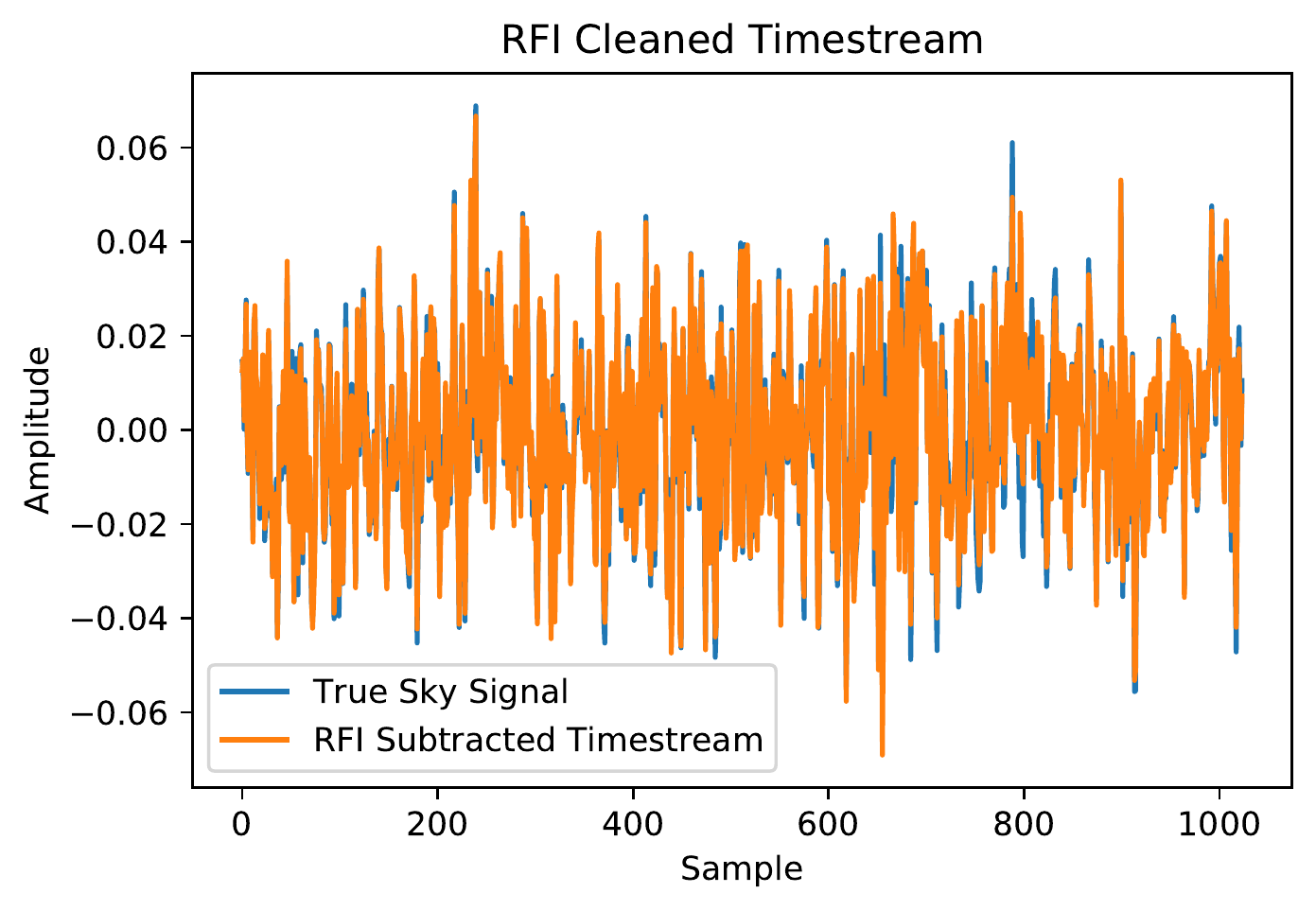}}
\subfigure{\includegraphics[width=0.49\textwidth]{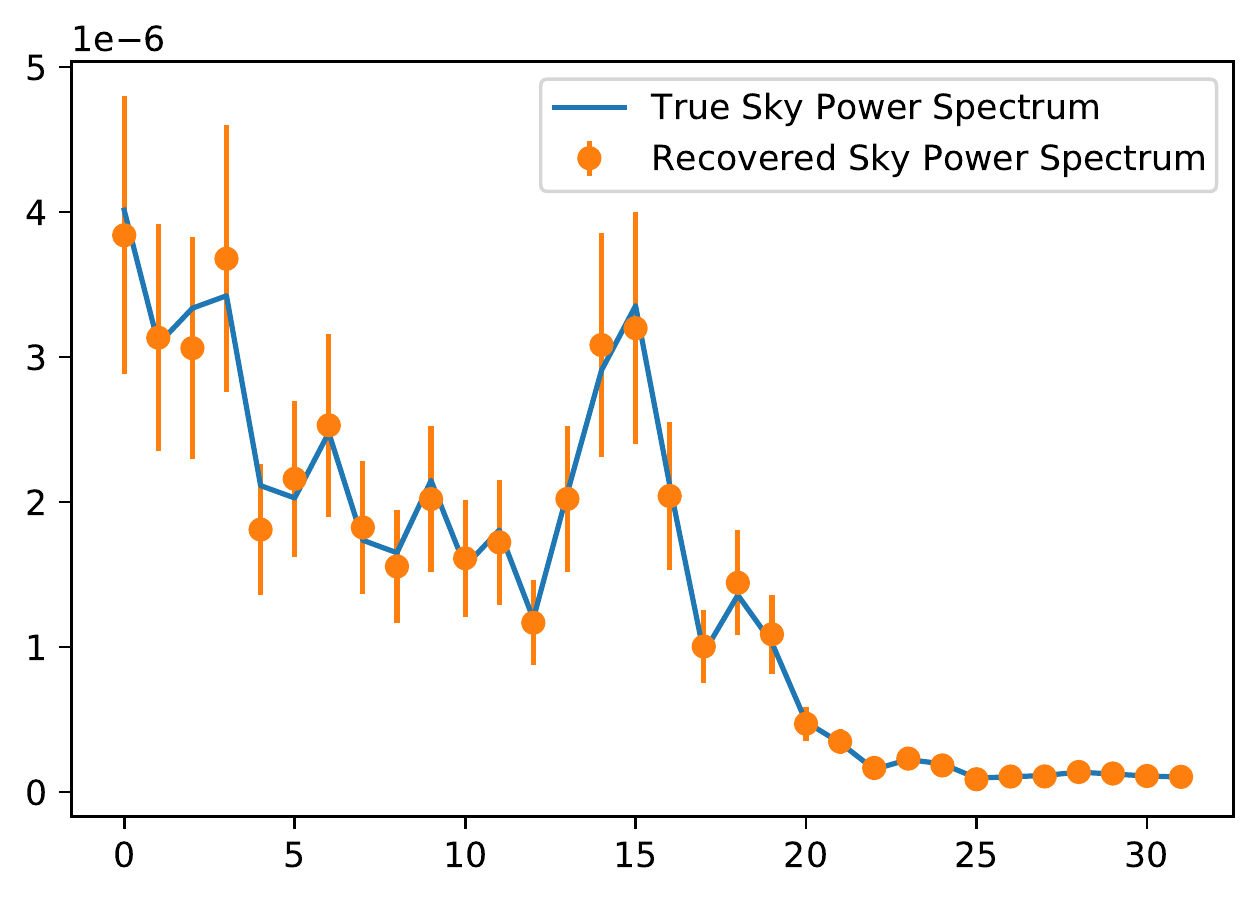}}
\end{center}
\caption[]{(Left) Recovered RFI-subtracted timestream compared to the input sky signal for the same timestream shown in Figure \ref{fig:rfi_overplot}. Horizontal axis is sample number, vertical is arbitrary amplitude units. Only minor errors in the recovered timestream can be seen, after subtracting an RFI event with peak amplitude nearly an order of magnitude larger than the sky signal. On average, the input and recovered timestreams differ by $0.56\%$ in amplitude in our tests. These errors do not significantly affect the recovered power spectra. (Right) Input and recovered sky power spectra for the timestream shown on the left. The horizontal axis is frequency and vertical is power, both in arbitrary units. Error bars are $1\sigma$ uncertainties, showing the input power spectrum is recovered to well within the uncertainty. The $\chi^2$ value for this recovered timestream is 0.0496, with 31 degrees of freedom, for a p-value of 1 to within  within $3 \times 10^{-38}$.}
\label{fig:pwr_spec} 
\end{figure}

We also want to verify that our RFI removal process does not substantially alter the recovered sky signal, or bias our estimation of the power spectrum of the sky signal. The left plot in Figure \ref{fig:pwr_spec} shows the same timestream as Figure \ref{fig:rfi_overplot}, but compares the input sky signal to the recovered timestream after subtracting RFI. The recovered timestream is almost identical to the input timestream, even after subtracting an RFI event with peak amplitude nearly an order of magnitude higher than the underlying sky signal. On average, the input and recovered timestreams differ by $0.56\%$ in amplitude. The errors that do occur can be seen mostly around the location of the RFI event. In Figure \ref{fig:pwr_spec}, this is from roughly time index 650 to 800.

The input timestreams seem to be recovered very well. However, it is possible that the recovered power spectrum may be biased in a way that is not obvious by eye in the timestreams. To evaluate whether the sky power spectrum is accurately recovered, we compute the power spectrum for our input simulated sky signal, not including the RFI event, and for our recovered sky signal, after subtracting the RFI event with our NN. These two power spectra are compared in the right hand plot in Figure \ref{fig:pwr_spec} for the same timestream as before. Here we use 32 frequency bins for visibility, instead of the 513 bins in our native FFT. The error bars are the $1\sigma$ uncertainties on our bins, calculated as:

\begin{equation}
\sigma_{P(k)} = \sqrt{\frac{1}{2 N_{k}}} P(k)
\end{equation}

where the factor of 2 accounts for the real and complex degrees of freedom at each frequency, $N_k$ is the number of frequencies per bin, and $P(k)$ is the value of the $k^{\mathrm{th}}$ frequency bin in the power spectrum. The recovered power spectrum clearly agrees with the power spectrum of the input sky signal to within the estimated uncertainty for a single 1024 sample time-stream frame. However, one might expect that averaged over many frames the RFI subtraction could have systematic effects.

\begin{figure}[H]
\begin{center}
\subfigure{\includegraphics[width=0.49\textwidth]{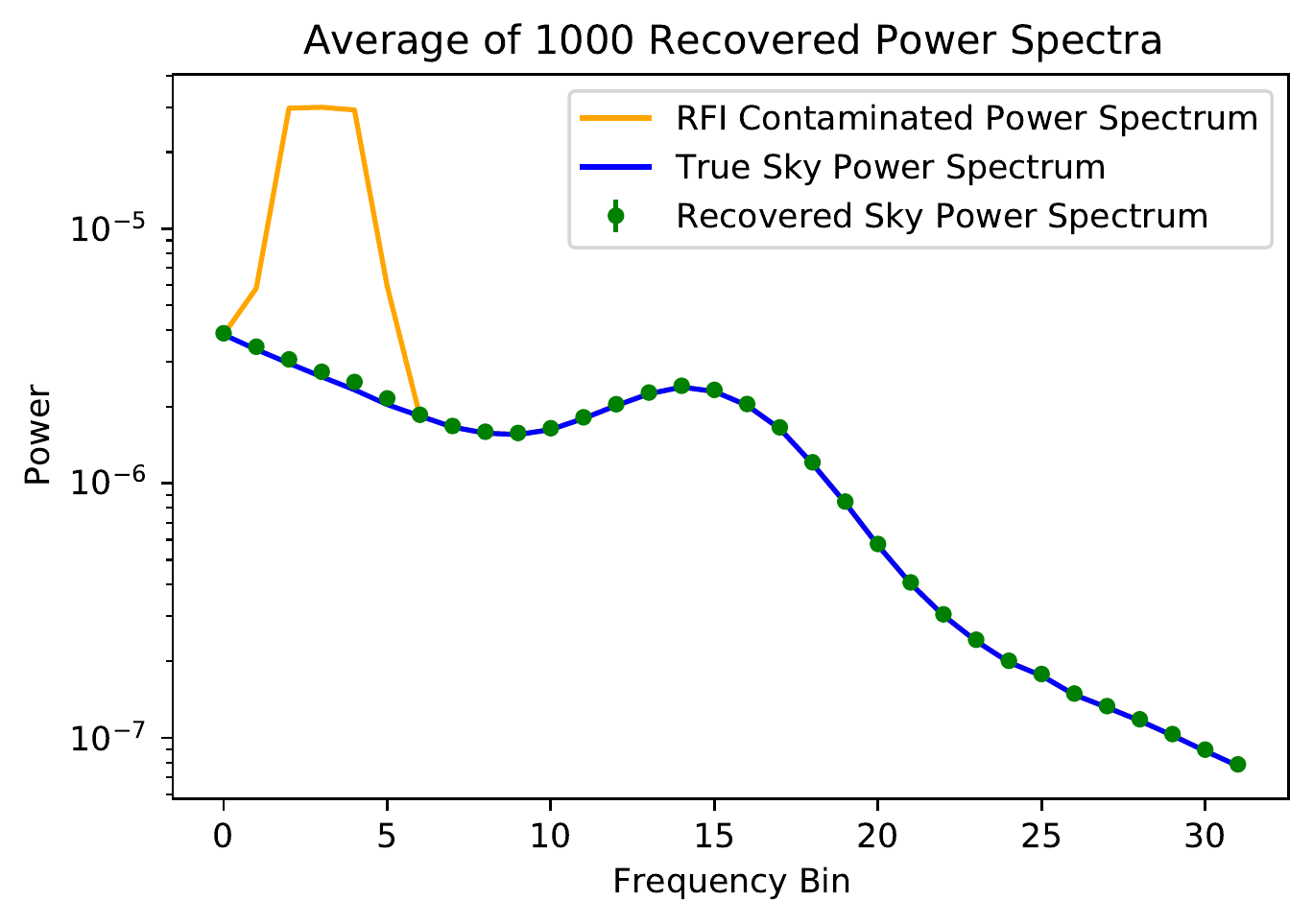}}
\subfigure{\includegraphics[width=0.49\textwidth]{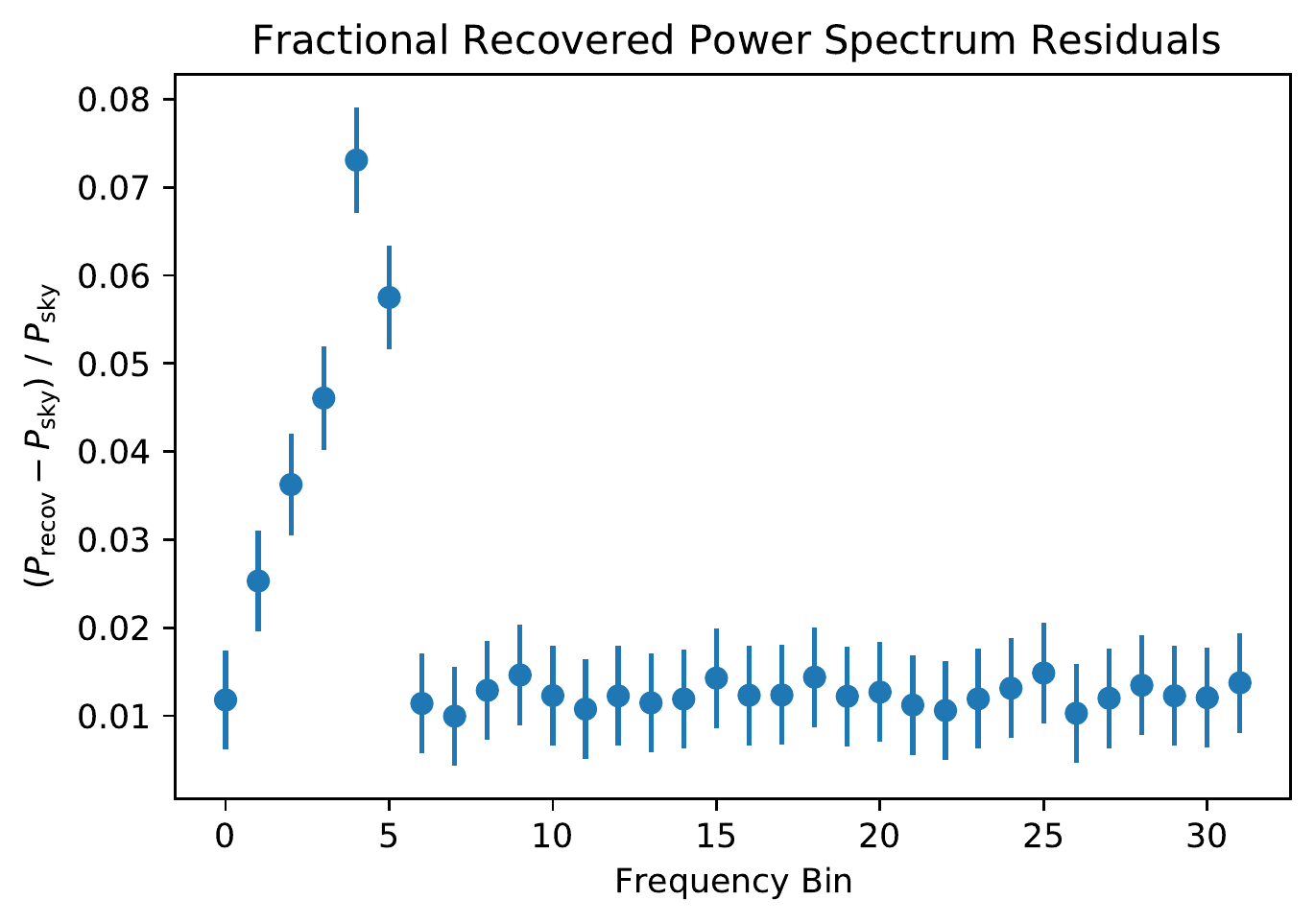}}
\end{center}
\caption[]{(Left) Input RFI Contaminated Power Spectrum (orange), true sky power spectrum (blue), and average recovered power spectrum from 1,000 timestream frames of length 1024 samples (green), showing the true sky power spectrum is consistently recovered despite RFI with peak power an order of magnitude greater than the underlying sky signal. $1\sigma$ error bars are smaller than data markers. (Right) Fractional residual power in each power spectrum bin. We find a $1\%$ increase in system noise across the band as a result of this RFI removal process, and a bias of up to $7\%$ in regions that were RFI contaminated. However, this bias is likely able to be characterized and calibrated, since it essentially acts as an additional noise.}
\label{fig:pwr_spec_n_1000} 
\end{figure}

To examine this possibility, we looked at 1,000 test timestreams using our fiducial NN, and localized RFI events as described in Section \ref{sec:optim}. We calculate the power spectra for the input RFI contaminated timestream, the true sky signal timestream, and the output RFI cleaned timestream produced by the NN. The left plot in Figure \ref{fig:pwr_spec_n_1000} shows the results. $1\sigma$ uncertainties in the recovered spectrum are plotted, but are smaller than the data points. The recovered power spectrum clearly follows the true sky spectrum, despite the RFI contaminated spectrum having an order of magnitude more power in frequency bins where it is present. 

To examine the bias produced by our NN, we plot the fractional residual power in each power spectrum bin, $(P_\mathrm{recov}-P_\mathrm{sky}) \ / \ P_\mathrm{sky}$, as show in the right plot in Figure \ref{fig:pwr_spec_n_1000}. We can see that the RFI removal process results in a $1\%$ increase in noise across the band, and an additional bias of up to $7\%$ of the sky power in RFI contaminated regions. 
This is also the bias for a maximally pessimistic case in which every frame is RFI contaminated. In reality, only $0.1\%$ to $1\%$ of frames are typically contaminated by RFI, taking BMX telescope data as our reference. A potential complication is that the amount of additional noise depends on the overall signal power in a non-linear way and thus it should be studied more carefully in a concrete application.  We will leave this calibration as a subject for future work. Significantly however, we have demonstrated that a data channel which is corrupted by RFI with amplitude an order of magnitude greater than the signal, and which would ordinarily have to be cut entirely, can be recovered with only a bias to the underlying sky power spectrum on the order of several percent in the RFI band.

\section{RFI Detection in BMX Data}
\label{sec:bmx}

The Baryon Mapping eXperiment (BMX) is an interferometric radio array at Brookhaven National Laboratory, which is designed as a pathfinder for future 21~cm intensity mapping experiments, including the proposed PUMA array \cite{slosar19}. It consist of four 4-m diameter off-axis parabolic dishes, with pyramidal feed horns and quad-ridge orthomode transducers. The signal chain includes temperature stabilized amplification and filtering, and a pulsed noise injection diode. It uses a third Nyquist-zone undersampled method to read out a 1.1-1.65~GHz observing bandwidth at 1.1~Gsamples/sec, at 8-bit precision. This observing bandwidth corresponds to a redshift range of $0 < z < 0.3$. A GPU implemented FX correlator produces 28~GB/day of time-ordered visibility data. Further details on the instrument design and calibration can be found in O'Connor et al. \cite{oconnor20}.

Here we use BMX time-ordered data obtained on November 7th, 2020 to train and test our NN. Approximately 2s of raw data were stored. This corresponds to $2^{31}$ samples for each of the 8 channels, which results in 16 GB of data. This data was converted into 1024 baseband channels of 537.1kHz in width and packetized in frames of 2048 samples, which then correspond to about 2ms.

We train on 100k timestreams from a single frequency channel, and evaluate 1k, using the optimized NN parameters from Section \ref{sec:optim}. Training takes 1040s, and evaluation takes 0.05s for the full 1k timestreams evaluated, or approximately 50$\mu$s per 2ms timestream (for a single frequency channel). At this rate, it would take approximately 25 cores to perform RFI filtering on baseband BMX data in real time, for a single polarization of a single dish. This underscores the necessity of implementing our NN methods in firmware to achieve the speeds required to be feasible for a large radio telescope array. 

\begin{figure}[H]
\begin{center}
\includegraphics[width=0.9\textwidth]{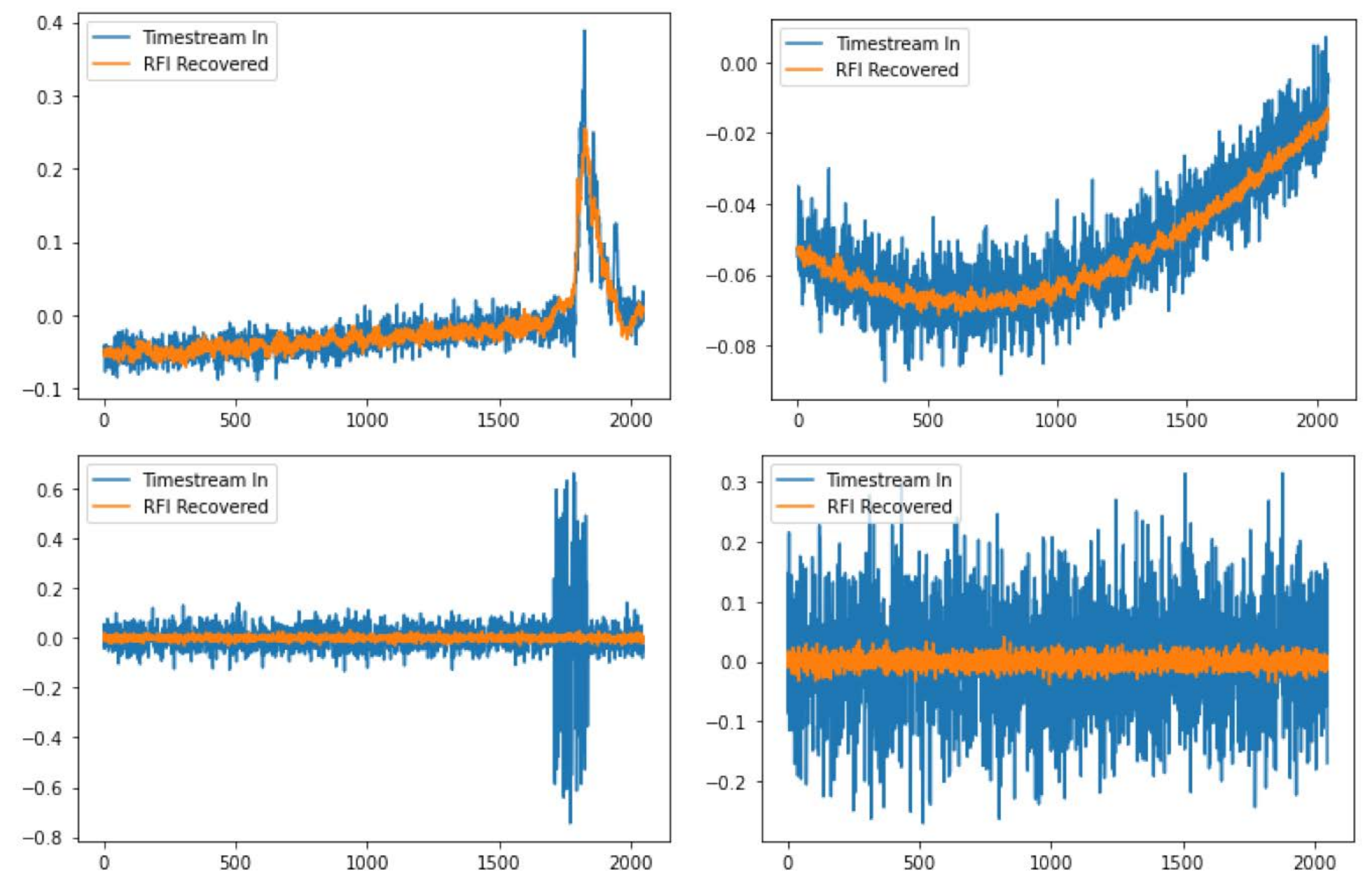}
\end{center}
\caption[]{Examples of detected RFI morphologies from BMX observations. Each timestream is 2048 samples long, for a total duration of 2ms. Vertical axis is arbitrary units. Top left: decaying pulse, Top right: long period sinusoidal event, Bottom left: broadband noise burst, Bottom right: null case for comparison. Our NN method is $>90\%$ efficient at detecting events of the pulse type, dependent on S/N, $>99.99\%$ efficient at detecting sinusoidal events, and does not detect broadband noise. In practice, broadband noise events would be flagged by a simple amplitude thresholding method run in parallel to the NN.}
\label{fig:bmx_rfi} 
\end{figure}

We have intentionally used the lowest frequency channels, which correspond to 1.1GHz, but are seen by the AC coupled digitizers as very low frequencies, and which have known contamination.  Figure \ref{fig:bmx_rfi} shows a sample of several timestreams, demonstrating some of the RFI morphologies present in the observed data, and the NN's ability to detect them. There are broadly three RFI morphologies present in our data sample, (1) spikes that rise quickly and decay more slowly in amplitude, (2) long period sinusoidal drift, and (3) broadband noise bursts. A sample of a clean timestream is also shown for comparison. We note that the long period sinusoidal event may not actually be RFI, but some aliasing effect in our edge frequency bins. This type of event is only detected in the edge frequency bins. However, it's detection does demonstrate the capability of the NN to detect and characterise such features in telescope timestreams. To assess the ability of the NN to detect RFI events we report false negative and false positive rates. However, since this is no longer simulated data but real sky data, we do not have access to the true underlying RFI. Therefore, in order to assess the NN's performance we visually inspected the 1,000 evaluation timestreams for RFI events and compared to the NN's output. This comparison is of course limited to RFI events which can be detected by eye, but serves as a baseline measurement of the NN's performance. Among the various RFI morphologies present, we find that the NN is most successful at detecting long period sinusoidal drift, with a false negative rate of $<$0.1\%. For the pulsed RFI morphology the detection rate is highly dependent on the signal to noise ratio of the event. Among RFI pulses that could be identified by eye in the sample timestreams evaluated, the NN had a false negative rate of 9.8\%. Our NN architecture proved unable to detect broadband noise pulses.  They can however easily be detected by an amplitude thresholding method, as described in Section \ref{sec:intro}. In practice, a thresholding algorithm would be used in parallel with the NN to detect broadband noise events, which would be relatively simple to implement. Reassuringly, the NN has an extremely low false positive rate. No false detections were seen in over 10k timestreams evaluated, or a $<$0.001\% false positive rate.

We note that these results were obtained used the neural network architechture as devised and tested on our toy model. Different settings might be better at identifying other features, such as the pulse type RFI seen.

\section{Discussion \& Conclusions}

In this paper we have developed a novel method for removing RFI signals from raw data streams. We have demonstrated that on a toy problem, the method can identify and successfully remove unwanted injected signals without ever being explicitly trained for a particular RFI component. It does so by finding compressible patterns in the data stream, which indicate a non-Gaussian signal and therefore a non-thermal origin.

This neural network method has the potential to retain more information than most RFI rejection methods, because it operates on the raw datastream, before any information loss. This also renders it computationally very intensive and would have to be implemented in low-level telescope firmware before being operational. It therefore cannot be be applied on current-generation telescopes, but could be applicable to future instruments. 

Modern hardware architectures naturally enable such levels of processing. For example in Tridgell et al.\cite{tridgell20} it has been demonstrated that the Xilinx ZCU111 RFSoC platform can evalute neural networks in real time. One could therefore imagine future radio interferometer architecture in which there is a per-feed channelizer that also implements neural-network based RFI rejection. In such a scheme, the neural network architecture would be periodically trained offline on a sample of raw waveforms with resulting weights transmitted to the individual elements for application to data. More futuristically, it might also be possible to simultaneously take data and use it to perform both operations, removing RFI with a current iteration neural network as well as simultaneously training an updated neural network for RFI removal in the future. To avoid noise-fitting it is of course not desirable to train and apply a network on the same data, but presumably the morphologies and population statistics of RFI sources affecting any given observatory will be constant or only slowly changing in time. In this case, the neural networks would enable self-learning instrumentation that can dynamically adapt to a changing RFI environment.

We note that while the algorithm is quite general as demonstrated by our blind application to BMX data, the sensitivity to different sources of RFI does depend on details of parameters of network architecture. In practice any deployment of such an approach to a real telescope would likely involve iterative training and testing and optimization of the neural network architecture.

In our work we have applied the network to the real raw data stream coming from the instrument ADC. This might not be an optimal choice. For example, it might be better to first process the data through a polyphase filter bank (PFB) channelizer and then apply RFI removal, before being further correlated and averaged. PFB samples contain the full information, but it has now been ``sorted" into the frequency domain while retaining time information.  For typical RFI sources, which are localized both in time and in frequency, this might be a better space. One could even imagine further processing PFB results into a sky-localized representation of the electric field. For example, for interferometers employing an FFT correlator,  the input channelized signal is first Fourier transformed to a real-space map of the electric field. The RFI hunting neural network could be applied to pixels of such a map before squaring it into an intensity map.  This might also enable choosing an optimal ratio of bandwidth and time over which the RFI filter processes data. For example, at 1GS/s, a 1024 sample window covers only approximately 1\,$\mu$s of data, far shorter than a typical RFI pulse. One could work on longer time windows, but the resource utilization might also grow. Instead, one could work (in parallel) over 10MHz windows with 100\,$\mu$s data chunks. We leave such optimizations for future work with applications to real data.

% References
\bibliography{saliwanchik} % bibliography data in report.bib
\bibliographystyle{spiebib} % makes bibtex use spiebib.bst

\end{document}